\documentclass[aps,prd,twocolumn,groupedaddress,showpacs,showkeys]{revtex4}

\usepackage{graphicx}
\usepackage{amssymb}
\usepackage{amsmath}

\usepackage{color}

\newcommand{\del}[1]{\textcolor{magenta}{\tiny #1}}
\newcommand{\add}[1]{\textcolor{red}{ #1}}

\begin{document}

\newcommand{\be}{\begin{equation}}
\newcommand{\ee}{\end{equation}}

\title{Spectroscopic Probe of the van der Waals Interaction\\ between Polar Molecules and a Curved Surface}

\date{\today}

\author{Giuseppe Bimonte$^{1,2}$, Thorsten Emig$^{3,4,5}$, R. L. Jaffe$^{5,6}$, and Mehran Kardar$^5$}

\affiliation{$^1$Dipartimento di Fisica E. Pancini, Universit\`{a} di
Napoli Federico II, Complesso Universitario
di Monte S. Angelo,  Via Cintia, I-80126 Napoli, Italy}
\affiliation{$^2$INFN Sezione di Napoli, I-80126 Napoli, Italy}
\affiliation{$^3$ LPTMS, CNRS, Universit\'e Paris-Saclay, 91405 Orsay, France}
\affiliation{$^4$Massachusetts Institute of Technology, MultiScale Materials Science
for Energy and Environment, Joint MIT-CNRS Laboratory (UMI 3466),
Cambridge, Massachusetts 02139, USA}
\affiliation{$^5$Massachusetts Institute of
Technology, Department of Physics, Cambridge, Massachusetts 02139, USA}
\affiliation{$^6$Center for Theoretical Physics, Laboratory for Nuclear Science, Massachusetts Institute of
Technology, Cambridge, Massachusetts 02139, USA}

\begin{abstract}
We  study the shift of rotational levels of a diatomic polar molecule due to its van der Waals (vdW) interaction with a gently curved dielectric surface at temperature $T$, and submicron separations. The molecule is assumed to be in its electronic and vibrational ground state, and the rotational degrees are described by a rigid rotor model. We show that under these conditions retardation effects and surface dispersion can be neglected. The level shifts are found to be independent of $T$, and given by the quantum state averaged classical electrostatic interaction of the dipole with its image on the surface.  We use a derivative expansion for the static Green's function to express the shifts in terms of surface curvature. 
We argue that the curvature induced line splitting is experimentally 
observable, and not obscured by natural line widths and thermal broadening.
\end{abstract}

\maketitle

\section{Introduction}

The van der Waals (vdW) interaction of neutral particles like atoms and molecules with macroscopic surfaces underlies many surface induced processes in physics, chemistry and biology~\cite{parse}. 
Also appearing in the guises of London and Casimir-Polder forces~\cite{london,polder}
these interactions originate from quantum dipole fluctuations of the particle that induce correlated fluctuations on the surface. 
While generally attractive, resonant coupling to surface excitations can lead to  repulsive forces~\cite{failache}. 
These fluctuation induced forces have typically been measured  for macroscopic bodies, 
while the vdW interaction of a free atom or molecule is less studied. 

Vacuum fluctuations of the electromagnetic field not only  give rise to Casimir forces between bodies,
but also have   observable effects on {\it isolated} particles, 
notably  they modify  energy levels of an atom,  an effect known as the Lamb shift.  
When a quantum particle is brought near a surface, the vdW interaction perturbs its energy levels. 
It has been shown that surface curvature leads to small corrections to the interaction of the particle with the 
surface~\cite{bimo1,bimo2}. Hence, one can expect also small corrections to the level shifts due to curvature. 
Here we shall demonstrate and explicitly quantify these shifts for the rotational levels of polar molecules.

For a flat metallic surface, the attractive vdW interaction potential was first measured with high precision for a sodium atom 
in 1992 by looking at the shifts of spectral lines using laser spectroscopy in the micrometer distance range.~\cite{haroche}  
More recently, for a sapphire surface supporting polariton excitations, a repulsive vdW potential acting on excited cesium atoms 
was observed in the $~100$nm distance range, by using selective reflection spectroscopy that allows for the observation of short-lived states~\cite{failache}.  Thermal  fluctuations within a hot surface can excite surface-polariton modes which can cause a strong temperature dependence of the vdW interaction.  Indeed, an up to 50$\%$ increase was measured spectroscopically for a cesium atom at short distances of $~100$nm away from a sapphire surface in the $500$ to $1000$K temperature range~\cite{ducloy}.

Unlike atoms, polar molecules have rotational and vibrational states that can be excited by radiation, or via the interaction with fluctuations in macroscopic bodies. The corresponding transition energies are often small compared to thermal energies.  
The resulting rotational and vibrational heating of cold diatomic molecules placed near a hot surface can imposes severe lifetime limits to the trapping of these particles which is relevant to the development of `molecular chips' using structured surfaces~\cite{Buhmann2008}.  
These and other specially designed nano- or micro-structured surfaces provide another tool to control vdW interactions. Hence, it is important to understand the influence of a non-trivial surface geometries on the internal dynamics of polar molecules which is governed by their spectral transitions. Recently, the non-equilibrium vdW force on a polar molecule near a metallic surface 
was computed and shown to saturate for high temperatures, making it distinct from the interaction for atoms~\cite{Ellingsen2009}.

The paper is organized as follows: In the next section we review the general theory for the finite temperature Casimir--Polder interaction between a quantum particle in an excited state and a dielectric surface. In Sec. III we compute that shifts of the rotational levels of a diatomic molecule  in terms of the static Green's function, and summarize characteristic parameters for experimentally relevant molecules and surface materials. A derivative expansion for the Green's function of curved surfaces is presented in Sec. IV, and this result is then used in Sec. V to estimate the curvature corrections to the energy levels of a simple rigid rotor model for a diatomic polar molecule. 
Finally, in the last section the magnitude and curvature dependence of the transition lines  of the modified rotational spectrum is estimated, and their experimental observability is discussed.

\section{Casimir-Polder interaction: general formulae}

We consider a quantum particle in a non-degenerate state $|a \rangle$, placed at a point ${\bf r}$ having (minimum) distance $d$ from a dielectric surface $S$ at temperature $T$ (see Fig.1).  We assume the separation  $d$ be much larger than the particle's size,  such that the particle can be modeled as a  dipole.  The material constituting the surface is assumed to be homogeneous and isotropic, described by  (complex)  dynamic permittivity $\epsilon(\omega)$. 
The Casimir-Polder (CP) interaction of the particle with the surface engenders a shift $\Delta F_a$ in the  free energy of state $|a \rangle$.  
As shown in Refs.~\cite{wylie, laliotis}, $\Delta F_a$ can be conveniently expressed as a sum of two terms
\begin{figure}
\includegraphics [width=.9\columnwidth]{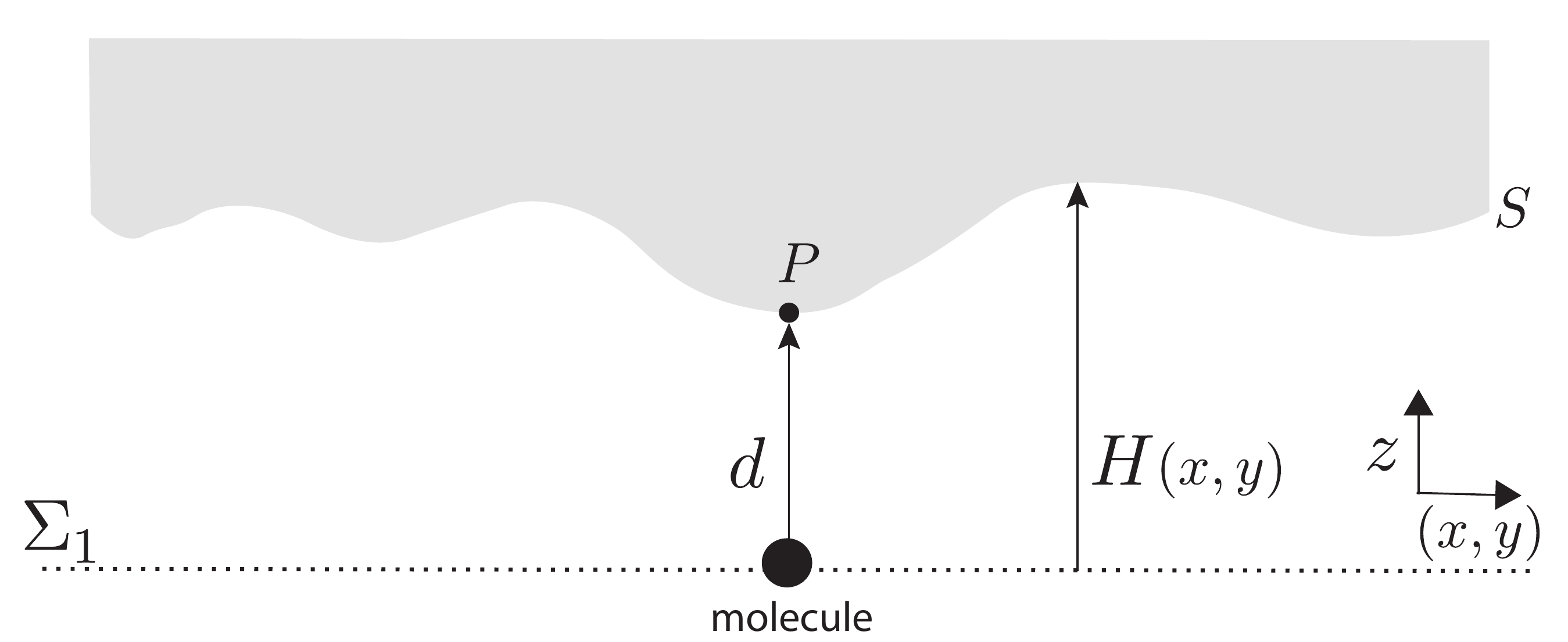}
\caption{\label{fig1} Parametrization of the profile of a gently curved dielectric surface near
an atom or molecule.}
\end{figure}
\be
\Delta F_a=\Delta F_a^{\rm nr}+\Delta F_a^{\rm r}\,.\label{CP}
\ee
The first term, $\Delta F_a^{\rm nr}$, 
is a non-resonant contribution having a form  similar to the familiar expression of the CP energy shift for a particle {\it in equilibrium} with a surface at temperature $T$~\cite{dalvit}:   
\be
\Delta F_a^{\rm nr}=- {k_B T}  \sum_{n=0}^{\infty\;'} \,\alpha_{ij}^{(a)} ({\rm i}\,\xi_n)  \;G^{(S)}_{ij}({\bf r},{\bf r}; {\rm i}\, \xi_n)\;,\label{CPeq}
\ee
while the second term represents a {\it resonant out-of-equilibrium} contribution:
\be
\Delta F_a^{\rm r}=\sum_{b \neq a} n(\omega_{ab},T)  \mu_i^{ab} \,\mu_j^{ba} {\rm Re}[G^{(S)}_{ij}({\bf r},{\bf r}; |\omega_{ba}|)]\;.\label{CPres}
\ee
In these equations $\omega_{ab}=(E_b-E_a)/ \hbar$ are the particle's transition frequencies, $\xi_n = 2 \pi n k_B T/\hbar$ are the (imaginary) Matsubara frequencies,   $\mu_i^{ba}=\langle b | \hat{\mu}_i | a \rangle$ are the matrix elements of the cartesian components (labelled by the latin index $i$) of the dipole moment  operator ${\hat{\bf {\mu}}}$, $n(\omega,T)=[\exp (\hbar \omega/k_B T)-1]^{-1}$ is the Bose-Einstein distribution function, the prime symbol in the sum over $n$ in Eq.~(\ref{CPeq}) indicates that the $n=0$ term is taken with weight 1/2, and $\alpha_{ij}^{(a)} ({\rm i}\,\xi)$ is the polarizability (relative to the state $|a \rangle$) of the particle:  
\be
\alpha_{ij}^{(a)} ({\rm i}\,\xi_n)=\frac{2}{\hbar}\sum_{b \neq a}  \mu_i^{ab} \,\mu_j^{ba} \frac{\omega_{ab}}{\xi_n^2+\omega_{ab}^2}\,.\label{polariz}
\ee 
Finally,  $G^{(S)}_{ij} ({\bf r},{\bf r}', \omega)$ denotes the (Fourier transform of the) surface contribution to the  electromagnetic Green's function, which is constructed as follows. 
Recall that the Green's function    $G_{ij} ({\bf r},{\bf r}', \omega)$   provides the electric field ${\bf E}({\bf r})$ at point ${\bf r}$ sourced by an oscillating dipole ${\bf p}(\omega)={\bf p}_0 e^{-{\rm i} \omega t}$   placed at the point ${\bf r}'$, as
\be
{E}_i({\bf r})=G_{ij} ({\bf r},{\bf r}',\omega)\, p_j (\omega)\,\del{,}\add{.}
\ee
The surface Green's function $G^{(S)}_{ij} ({\bf r},{\bf r}', \omega)$ is defined by  the following decomposition of  $G_{ij} ({\bf r},{\bf r}',\omega)$,
\be
G_{ij} ({\bf r},{\bf r}',\omega)=G^{(0)}_{ij} ({\bf r},{\bf r}',\omega)+G^{(S)}_{ij} ({\bf r},{\bf r}',\omega)\,,
\ee
where $G^{(0)}_{ij} ({\bf r},{\bf r}',\omega)$ is the free-space Green's function.  Thus $G^{(S)}_{ij} ({\bf r},{\bf r}',\omega)$ can be physically interpreted as describing the field generated by the induced dipoles   on the surface $S$. 
We note that in the coincidence limit ${\bf r}={\bf r}'$, the surface Green's function $G^{(S)}_{ij} ({\bf r},{\bf r}',\omega)$ attains a finite limit (unlike from the free space contribution $G^{(0)}_{ij} ({\bf r},{\bf r}',\omega)$ which diverges in this limit), which ensures that the CP energy shift in Eq.~(\ref{CP}) is well defined. It is also important to bear in mind that the frequency-dependence of the surface Green's function $G^{(S)}_{ij} ({\bf r},{\bf r}',\omega)$ is twofold: besides an explicit frequency-dependence, due to {\it retardation effects},
there is the implicit frequency-depence due to {\it dispersion} in the response function $\epsilon(\omega)$  of  the surface. 

\section{Shifts of rotational levels of diatomic molecules}

We shall use  Eqs.~(\ref{CP}--\ref{CPres}) to estimate the   shifts $\Delta F_a$ of the rotational levels of  a polar diatomic molecule with   a closed electronic shell (i.e. in a $^1 \Sigma^+$ state), in its ground electronic and vibrational state (for a review of rotational spectroscopy of diatomic molecules see Ref.~\cite{brown}). 
 
Some characteristic parameters (the angular frequency $\omega_r$ and the wavelength $\lambda_r$  corresponding to transitions from the ground state to the first excited rotational state,  and the dipole moment $\mu$) of typical polar molecules are listed in Table I.   The computation of the shifts of rotational levels of diatomic molecules is indeed very simple, thanks to the simplifying circumstance that in the evaluating Eqs.~(\ref{CP}--\ref{CPres}) both sources of frequency dependence in the dynamic Green's function $G^{(S)}_{ij} ({\bf r},{\bf r}',\omega)$, i.e. retardation effects and surface dispersion, can be 
neglected. 

Let us consider retardation effects first. We will see later on that measurable shifts of the rotational levels  occur only for  submicron separations between the molecule and the  surface. For such small separations, we can safely neglect  retardation effects.  This is so because for a polar diatomic molecule the largest matrix elements $\mu_i^{ba}$ of the dipole moment operator are relative to transitions between adjacent rotational levels~\cite{brown}, 
which have characteristic frequencies of order $\omega_r$. This implies at once that both the resonant and the non-resonant contributions to the shift $\Delta F_a$ are dominated by frequencies of order $\omega_r$ or smaller.   This is obvious for the resonant contribution $\Delta F_a^{\rm r}$, because from  Eq. (\ref{CPres}) we see that the frequency argument of $G^{(S)}({\bf r},{\bf r}; |\omega_{ba}|)$ is indeed of order $\omega_{r}$. As to the non resonant contribution, we see from Eq. (\ref{CPeq}) that $\Delta F_a^{\rm nr}$ receives its dominant contribution from the Matsubara modes $\xi_n$  such that the molecule's polarizability $\alpha_{ij}^{(a)} ({\rm i}\,\xi_n)$ is significant. In view of Eq. (\ref{polariz}) it is clear that this is the case only if $\xi_n$ is of order $\omega_r$ or smaller.   
It follows from these considerations that retardation effects become important only for separations  of the order of  $\lambda_r=2 \pi c/\omega_r$ or larger.  
As  seen from Table I, the wavelength of transitions between rotational states   of diatomic molecules  is of the order of millimeters, showing that for experimentally relevant distances retardation effect are indeed negligible.       

\begin{table}[h]
\begin{tabular}{l l l l} \hline
\; & $\omega_r (10^9 {\rm rad/s})$  & $\;\;\;\lambda_r $(mm) & \;\;\;$\mu (10^{-30} \,{\rm C \,m})$  \\ \hline \hline
LiH\;\;\; &2790 &  \;\;\;0.7 &\;\;\;  19.6\\ 
LiRb\;\;\;& 83 & \;\;\,22.7 &\;\;\; 13.5 \\
LiCs \;\;\;& 73&  \;\;\;25.8 &\;\;\;\;21.0 \\
NaRb\;\;\;\;& 25.5 &\;\;   73.8 &\;\;\;\;11.7 \\
NaCs\;\;\;&22.2 & \;\;\, 84.8 & \;\;\;\;19.5 \\ \hline
\end{tabular}
\caption{ Characteristic parameters of some polar diatomic molecules with closed electron shells.   }
\end{table}

Dispersion effects within the surface can also be ignored as the    
angular motion of diatomic molecules is much slower than relaxation processes characterizing typical dielectric materials.    
Common dielectrics used in  experiments are sapphire, ${\rm Ca F}_2$, ${\rm Ba F}_2$ and SiC. 
Among these, sapphire is frequently employed in atom-surface interaction experiments, 
while SiC is normally used in experiments on near-field heat transfer. 
The common feature of these materials is that their optical properties is well described by a single-resonance model over a wide range frequencies extending to  visible range. 
In  this model, the complex permittivity  $\epsilon(\omega)$ is described by 
\be
\epsilon(\omega)=\epsilon_{\rm inf}+\frac{(\epsilon_{\rm st}-\epsilon_{\rm inf})\,\omega_T^2}{\omega_T^2-\omega^2-{\rm i} \Gamma \omega}\;,\label{permit}
\ee
where  $\epsilon_{\rm st}$  and  $\epsilon_{\rm inf}$ represent the static and optical dielectric constants respectively, 
$\Gamma$ is a phenomenological relaxation frequency, and $\omega_T$ is  
the transverse optical (TO) phonon frequency.  Values of these parameters for the materials considered  are listed in Table II~\cite{laliotis}.
\begin{table}[h]
\begin{tabular}{l l l l l} \hline
\;\;\; & $\epsilon_{\rm st}$  &\;\;\;\;\; $\epsilon_{\rm inf}$ &\;\;\;  $\omega_T (10^{12} {\rm rad/s})$ & $\;\;\; \Gamma  (10^{12} {\rm rad/s})$  \\ \hline \hline
${\rm BaF}_2$ \;\; &7.16 & \;\;\;\;\;\;\,2.12 & \;\;\;\;\;\; 33.9 &\;\;\;\; 0.4 \\ 
${\rm CaF}_2$ &  6.82 & \;\;\;\;\;\; 2.02 & \;\;\;\;\;\; 48.7 &\;\;\;\;   0.8\\
 Sapphire  &    9.32& \;\;\;\;\;\; 3.03& \;\;\;\;\;\; 97.6 &\;\;\;\;  0.5\\
SiC &  10 & \;\;\;\;\;\; 6.7 & \;\;\;\;\;\;  149.4 &\;\;\;\;  0.14 \\\hline
\end{tabular}
\caption{Parameters for complex permittivity of sapphire, ${\rm Ca F}_2$, ${\rm Ba F}_2$ and SiC.}
\end{table}
According to Eq.~(\ref{permit}) the frequency dependent permittivity $\epsilon(\omega)$ is well approximated by the static dielectric constant $\epsilon_{\rm st}$ for frequencies $\omega \ll \omega_T$.  
The shifts $\Delta F_a$ of the rotational levels of a molecule  arise mostly from transitions between adjacent rotational states, with characteristic frequencies of the order of $\omega_r$. By comparing Table I with Table II, we see that for all considered molecules and dielectrics $\omega_r \ll \omega_T$, and thus the static permittivity $\epsilon_{\rm st}$ of the surface can be safely used to estimate the shifts $\Delta F_a$. 

Summarizing the above considerations, for experimentally relevant molecule-surface separations and for realistic dielectric materials, the CP energy shifts of rotational levels  of diatomic molecules can be estimated by substituting into Eqs.~(\ref{CP}--\ref{CPres}) the static Green' function ${{\bar G}^{(S)}_{ij}}({\bf r},{\bf r}; \epsilon_{\rm st})$ for 
the full dynamical Green's function   $G^{(S)}_{ij}({\bf r},{\bf r}; {\rm i}\, \xi_n)$  or $G^{(S)}_{ij}({\bf r},{\bf r}; {\rm i}\, |\omega_{ba}|)$. In what follows, we shall denote by ${\bar G}^{(S)}_{ij}(d,\epsilon_{\rm st}) \equiv {\bar G}^{(S)}_{ij}({\bf r}, {\bf r}, \epsilon_{\rm st}) $ the static Green's function of the surface $S$ evaluated at the position ${\bf r}$ occupied by the molecule.  After substituting $G^{(S)}_{ij}({\bf r},{\bf r}; {\rm i}\, \xi_n)$  by  ${\bar G}^{(S)}_{ij}(d,\epsilon_{\rm st})$, the expression for $\Delta F_a\add{^{\rm nr}}$ simplifies considerably. 
Summing over the Matsubara frequencies, $\Delta F_a^{\rm nr}$ is obtained as
\be
\Delta F_a^{\rm nr}=-\frac{1}{2} {{\bar G}^{(S)}_{ij}}(d, \epsilon_{\rm st}) \sum_{b \neq a} \mu_i^{ab} \,\mu_j^{ba} \;\coth \left(\frac{\hbar \omega_{a b}}{2 k_B T} \right)\;.\label{CPeqmol}
\ee
Similarly for $\Delta F_a^{\rm r}$,  using the identity
\be
n(\omega,T)=\frac{1}{2} \left[\coth \left(\frac{\hbar \omega_{a b}}{2 k_B T} \right)-1 \right]\;,
\ee
and noting that  $ {\rm Re}[{{\bar G}^{(S)}_{ij}}(d, \epsilon_{\rm st}) ]={{\bar G}^{(S)}_{ij}}(d, \epsilon_{\rm st})$ since $\epsilon_{\rm st}$ is real, we find
\be
\Delta F_a^{\rm r}=\frac{1}{2}{{\bar G}^{(S)}_{ij}}(d, \epsilon_{\rm st}) \sum_{b \neq a} \mu_i^{ab} \,\mu_j^{ba} \left[\coth \left(\frac{\hbar \omega_{a b}}{2 k_B T} \right)-1 \right]\;.\label{CPresmol}
\ee
Adding Eqs.~(\ref{CPeqmol}) and (\ref{CPresmol}), now leads to the compact form (see also Eq. (10) of Ref.\cite{Ellingsen2009})
\begin{align}
\Delta F_a&=&-\frac{1}{2}\;{{\bar G}^{(S)}_{ij}}(d, \epsilon_{\rm st}) \; \sum_{b \neq a} \mu_i^{ab} \,\mu_j^{ba} \nonumber \\
&=& -\frac{1}{2}\;{{\bar G}^{(S)}_{ij}}(d, \epsilon_{\rm st}) \;  \langle a \vert \mu_i \,\mu_j \vert a \rangle\;.\label{shiftfinal}
\end{align}
The final result is very simple: it shows that the energy shift of the rotational state $|a \rangle$ of a diatomic molecule is independent of the surface temperature, and coincides with the classical electrostatic  interaction energy of the dipole with its image on the  surface~\cite{jackson}, averaged over the quantum state $|a \rangle$ of the molecule.  The temperature independence of the non-retarded Casimir-Polder potential for a molecule placed near a dielectric surface has been noted before in the literature, as a result of cancellations between non-resonant potential components  and those due to evenescent waves \cite{Ellingsen2009,Ellingsen2010}.

\section{Derivative expansion of the static Green's function}

The static Green's function  ${\bar G}^{(S)}_{ij}(d,\epsilon_{\rm st})$ for a dielectric surface $S$, even if simpler than the dynamic Green's function $G^{(S)}_{ij} ({\bf r},{\bf r}',\omega)$, still cannot be determined for surfaces of arbitrary shapes. 
Analytical expressions for ${\bar G}^{(S)}_{ij} ({\bf r},{\bf r}', \epsilon)$
are known only for simple geometries of the surface  such as planes and spheres~\cite{smythe}, while for general shapes the problem has to be attacked numerically. 
Here we show that   a {\it derivative expansion} can  be used to obtain the asymptotic small-distance form of ${\bar G}^{(S)}_{ij}(d,\epsilon_{\rm st})$ for any gently curved dielectric surface.   
The derivative expansion has been recently applied successfully to estimate curvature corrections to the Casimir interaction between two gently curved 
surfaces~\cite{fosco2,bimonte3,bimonte4}, and to the CP interaction of a   nanoparticle with a curved surface~\cite{bimo1,bimo2}. 
Here, we apply it to the CP interaction of a quantum particle with a surface.

Let us denote by $\Sigma_1$ (see Fig.~1) a plane through the molecule which is orthogonal to the distance vector (which we take to be ${\hat {\bf z}}$ axis) connecting the molecule to the point $P$ of the surface $S$ closest to the molecule. We assume that the surface $S$ is described by a {\it smooth} profile $z=H({\bf x})$, where ${\bf x}=(x,y)$ is the vector spanning $\Sigma_1$, with origin at the molecule's position. In what follows latin indices $i,j,k \dots$ shall label all coordinates $(x,y,z)$, while greek indices $\alpha, \beta, \dots$ shall refer to coordinates $(x,y)$  on the plane $\Sigma_1$. 

In the present context, the key idea behind the gradient expansion is simple to explain: 
As dipole-dipole interaction falls off rapidly with distance, it is reasonable to expect that 
for small separations $d$ the Green's function ${\bar G}^{(S)}_{ij}(d)$ is mainly determined by the shape of the surface $S$ in a small 
neighborhood of the point $P$ closest to the molecule.
This physically plausible idea suggests that for small separations the Green's function can be expanded as a series  
in an increasing number of derivatives of the height profile, evaluated at the molecule's position. 
Up to second order, and assuming that the surface is homogeneous and isotropic, the most general expression  that is invariant under rotations of the $(x,y)$ coordinates, and that involves at most two derivatives of $H$
[but no first derivatives, since  $\boldsymbol{\nabla} H ({\bf 0})=0$]  has the form
\begin{align}
{\bar G}^{(S)}_{\alpha \beta}(d)&={\bar G}^{(\rm plane)}_{\alpha \beta}(d)+\frac{1}{32 \pi \epsilon_0 \, d^2} \left\{  \; \beta_2^{(2)} \nabla^2 H  \delta_{\alpha \beta} \right.  \nonumber\\
&\left.  +\;\beta_3^{(2)}\left(\partial_{\alpha} \partial_{\beta} H-\frac{1}{2} \delta_{\alpha \beta} \nabla^2 H \right)  \right\}\;,\label{gradexp2}
\end{align}
\begin{align}
{\bar G}^{(S)}_{zz}(d)&={\bar G}^{(\rm plane)}_{zz}(d)+\frac{\beta_1^{(2)}}{32 \pi \epsilon_0 \, d^2}    \nabla^2 H  \;,\label{gradexp1}\\
{\bar G}^{(S)}_{\alpha z}(d)&={\bar G}^{(S)}_{z\alpha}(d)=0\;.
\end{align}
Here,  $\boldsymbol{\nabla}$ is the gradient in the plane $\Sigma_{1}$, $\epsilon_0$ is the vacuum permittivity, ${\bar G}^{(\rm plane)}_{ij}(d)$ is the well known Green's function for a planar dielectric surface, 
while the coefficients $\beta^{(2)}_q$ are dimensionless functions of the permittivity $\epsilon$. The geometric significance of the expansion in   Eqs.~(\ref{gradexp2}--\ref{gradexp1}) becomes more transparent when $x$ and $y$ are chosen to be coincident with the principal directions of $S$ at $P$, in which case the local expansion of $H$ takes the simple from 
 $ H(x,y)=d + {x^2}/{(2 R_1)}+ {y^2}/{(2 R_2)}+ \dots$, where $R_1$ and $R_2$ are the radii of curvature at $P$. To be definite, we assume that 
${d}/{R_1} \ge {d}/{R_2}$. 
In this coordinate system, the derivative expansion of ${\bar G}^{(S)}_{ij}(d, \epsilon)$ reads:
\be
{\bar G}^{(S)}_{zz}(d)={\bar G}^{(\rm plane)}_{zz}(d)+\frac{\beta_1^{(2)} }{32 \pi \epsilon_0 \, d^3}   \left(\frac{d}{R_1}+\frac{d}{R_2} \right)  \;,\label{gradexp3}
\ee
\begin{align}
{\bar G}^{(S)}_{xx/yy}(d)&= {\bar G}^{(\rm plane)}_{xx/yy}(d)+\frac{1}{32 \pi \epsilon_0 \, d^3}  \left[   
 \beta_2^{(2)}   \left(\frac{d}{R_1}+\frac{d}{R_2} \right)   \right.
\nonumber \\
&\left. \pm\frac{\beta_3^{(2)}}{2}  \left(\frac{d}{R_1}-\frac{d}{R_2} \right)  \right]   \;.\label{gradexp4}
\end{align}
The procedure to determine the coefficients $\beta^{(2)}_q$  is explained in detail in Refs.~\cite{bimo1,bimo2}, and based on the following:
The derivative expansion in  Eqs.~(\ref{gradexp2}--\ref{gradexp1})  is valid for surfaces of small-slope, i.e. for $d/R \ll 1$ where $R$ is a characteristic radius of curvature. 
However, for  height profiles of small amplitude $H(x,y)=d+ h(x,y)$ such that $h(x,y)/d \ll 1$,  the Green's function ${\bar G}^{(S)}_{ij}(d)$ can  also be Taylor expanded in 
powers of  $h(x,y)$. It is sufficient to consider the latter expansion to first order in $h(x,y)$,
\be
{\bar G}^{(S)}_{ij}(d)={\bar G}^{(\rm plane)}_{ij}(d)+ \int \frac{d^2 {\bf k}}{(2 \pi^2)} {\bar G}^{(2)*}_{ij}( {\bf k},d) {\tilde h}({\bf k})\;,\label{taylor}
\ee 
where ${\bf k}$ is the in-plane wave vector and ${\tilde h}({\bf k})$ is the Fourier transform of the $h(x,y)$. 
After the kernel $ {\bar G}^{(2)}_{ij}( {\bf k},d)$ is  computed,  the coefficients
$\beta^{(2)}_q$ are determined by matching, in the common domain of validity,  the derivative expansion of ${\bar G}^{(S)}_{ij}(d)$ in Eqs.~(\ref{gradexp1}--\ref{gradexp2}) with the Taylor expansion in Eq.~(\ref{taylor}).  
By following these steps one arrives at the following small-distance expansion:
\begin{widetext}
\be
{{\bar G}^{(S)}_{xx/yy}}(d)= \frac{1}{32 \pi \epsilon_0 \, d^3} \frac{\epsilon-1}{\epsilon+1} \left\{1-  \frac{5 + 3 \epsilon}{4(\epsilon+1)} \left(\frac{d}{R_1}+\frac{d}{R_2} \right) \mp   \frac{1 + 3 \epsilon}{8(\epsilon+1)} \left(\frac{d}{R_1}-\frac{d}{R_2} \right) + O\left(({d\over R})^2\right)\right\}\; ,\label{Gxxyy}
\ee
\be
{{\bar G}^{(S)}_{zz}}(d)= \frac{1}{16 \pi \epsilon_0 \, d^3} \frac{\epsilon-1}{\epsilon+1} \left\{1-  \frac{3 + \epsilon}{4(\epsilon+1)} \left(\frac{d}{R_1}+\frac{d}{R_2} \right) +  O\left(({d\over R})^2\right)
\right\}\;. \label{Gzz}
\ee  
\end{widetext}

\section{A simple model: the rigid rotor}

In this Section we use Eq.~(\ref{shiftfinal}), together with Eqs.~(\ref{Gxxyy}--\ref{Gzz}), to estimate the shifts $\Delta F_a$ of the rotational levels of a diatomic polar molecule, near a gently curved surface. To estimate the matrix elements of the dipole-moment operator in the rotational states of the molecule in its ground electronic state, we shall model the  diatomic polar molecule as a simple rigid rotor~\cite{brown}.   In what follows, we shall neglect the hyperfine structure of the rotational spectrum. For  molecules in a $^1 \Sigma^+$ state the hyperfine structure is mainly due to the electric quadrupole interaction between the nuclear quadrupole moment and the electric-field gradient at the nucleus~\cite{brown}. The nuclear quadrupole hyperfine splitting in  $^1 \Sigma^+$ states typically   ranges from tens of kHz to  one or two hundred kHz. We shall see later on that the level splitting determined by the CP interaction can be as large as several MHz,  which justifies neglecting the hyperfine structure.  

According to the rigid rotor model, far from the surface, the Hamiltonian operator ${\hat H}$ describing the molecule is
\be
{\hat H}=\frac{{\hat {\bf L}}^2}{2\, I}\;,
\ee
where   ${\hat {\bf L}}$ is the rotational angular momentum, and $I$ is the moment of inertia.  The energy eigenstates  $|l,m \rangle$  
are labelled by the quantum numbers $l=0,1,2,\dots$ and $m$, with $-l \le m \le l$  corresponding, respectively, to the rotational angular momentum and  to its $z$-component ${\hat {L}_z} $ (we choose as $z$ axis the line connecting the molecule to the point $P$ of the surface $S$ closest to the molecule, see Fig.~1), such that
\be
{\hat {\bf L}}^2 |l,m \rangle= \hbar^2\,l\,(l+1)|l,m \rangle\;,
\ee     
\be
{\hat {L}_z} |l,m \rangle=\hbar\,m  |l,m \rangle\;.
\ee
Then, 
\be
{\hat H}  |l,m \rangle=E_l  |l,m \rangle\;,
\ee
where
\be
E_l = \frac{\hbar \, \omega_{r}}{2} \,l\,(l+1)\;,
\ee 
and we set $\omega_r=\hbar/I$. The level of energy $E_l$ consists of $2 l+1$ degenerate states, distinguished by the azimuthal quantum number $m$.

When the molecule is brought near the surface, the CP interaction perturbs its energy levels. To analyze the effect of the interaction with the surface, we consider that for a gently curved surface such that $d/ R \ll 1$,  curvature effects  are expected to cause a small correction to the perturbation determined by a planar surface.  This suggests to split  the computation of the energy shifts $\Delta F_a$ in two steps: in the first step, we study the planar problem, and then we consider how the energy levels for a planar surface are further modified by curvature effects.  As we shall see below,  this procedure has the advantage  that it allows us to use the theory of CP energy shifts for non degenerate quantum states, presented in Sec.~II.

\subsection{A planar surface}

For a planar surface (and more  generally for any axisymmetric surface) the  Green's function ${\bar G}^{(S)}_{ij}(d)$ is invariant under rotations around the ${\hat z}$-axis, and therefore the azimuthal label $m$ remains  a good quantum number  in the presence of the surface. The CP interaction does not mix states with different values of $m$, and   therefore  we can straightforwardly use
the results in Sec.~II to compute the shifts $\Delta F_{l,m}$. 
Using the relations: 
$$
\langle l,m | \hat{\mu}_x^2 | l, m \rangle=\langle l,m | \hat{\mu}_y^2 | l, m \rangle
$$
\be
= \mu^2 \;\frac{l (l+1)+m^2-1}{4 l (l+1)-3}\;,
\ee
and
\be
\langle l,m | \hat{\mu}_z^2 | l, m \rangle=\mu^2 \;\frac{2 l (l+1)-2 m^2-1}{4 l (l+1)-3}\;,
\ee
we find
\be
\Delta F^{(\rm plane)}_{l,m}=-{\cal E}\,\frac{3 l(l+1)-m^2-2}{4 l(l+1)-3} \,,\label{plane}
\ee
where
\be
{\cal E} =\frac{\mu^2}{32 \pi \epsilon_0 \, d^3}\;\frac{\epsilon_{\rm st}-1}{\epsilon_{\rm st}+1} \;.\label{scale}
\ee
According to Eq.~(\ref{plane}), the  CP interaction of the molecule with a plane splits the $(2l+1)$-fold degenerate level $E_l$ into $l$ distinct levels of energies $E^{\rm (plane)}_{l,|m|}=E_l+\Delta F^{(\rm plane)}_{l,m}$, labelled by the absolute value of the azymuthal quantum number  $|m|$. Of these levels, only  $m=0$ is non degenerate, while those with $m \neq 0$ form degenerate doublets  (see Fig.~\ref{levelsS}). \\
 
\subsection{Curvature corrections}

\begin{figure}
\includegraphics [width=.9\columnwidth]{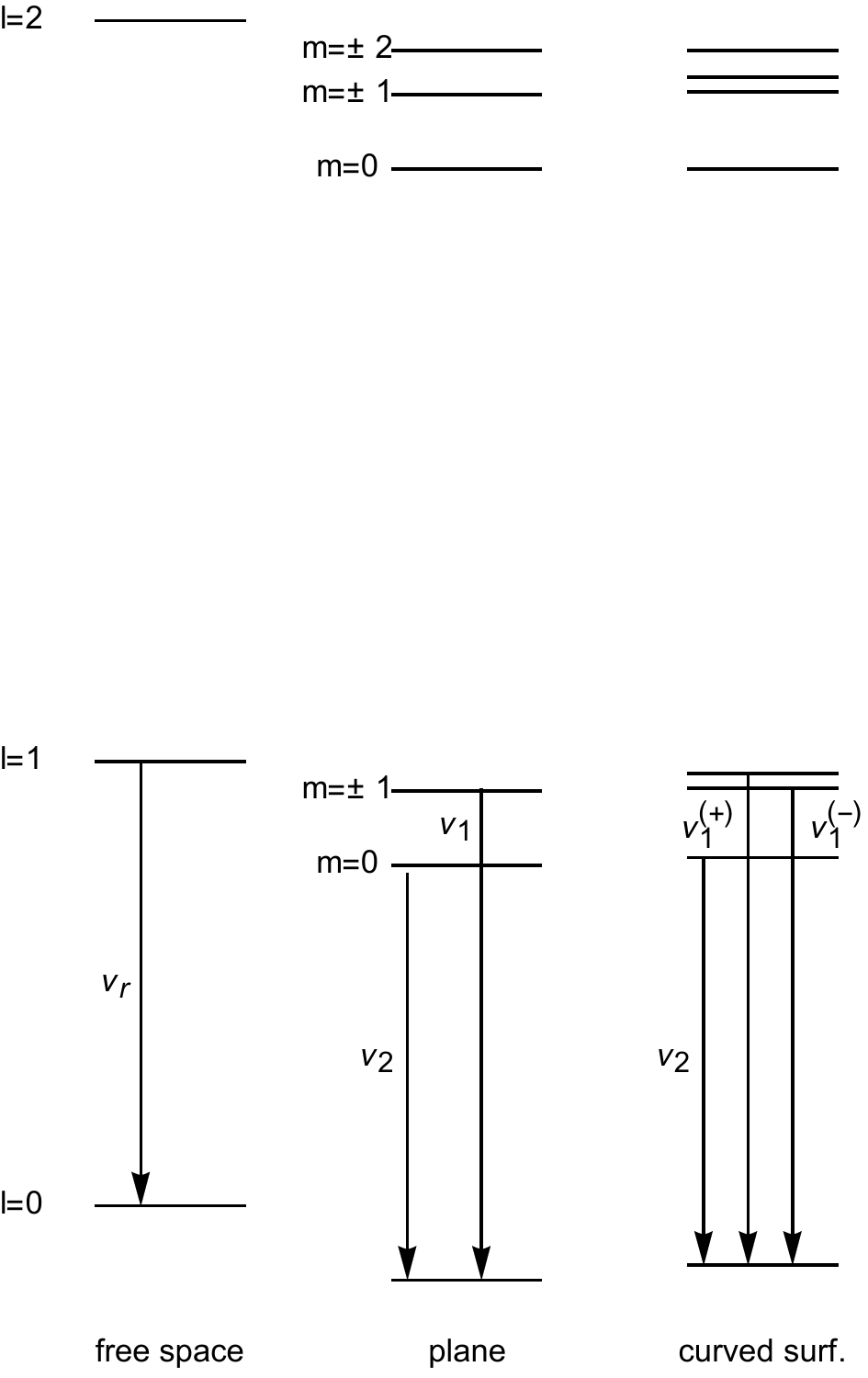}
\caption{\label{levelsS}  Qualitative structure of the energy levels of a diatomic polar molecule in free space (left),  near a planar surface (middle) and near a curved surface with different radii of curvature (right).}
\end{figure}

Having determined the structure of the energy levels $E_{l,|m|}^{(\rm plane)}$ of a diatomic molecule near a planar surface, we  now study how  the levels $E_{l,|m|}^{(\rm plane)}$  are affected  by the surface curvature.   As we pointed out above, we consider that  for $d/R \ll 1$ curvature corrections are small, compared to   the CP energy shift  for a planar surface. This suggests that  we  take the (possibly) doubly-degenerate levels  $E^{\rm (plane)}_{l,|m|}$ determined in the previous step as our unperturbed states, and compute  curvature corrections  to their energies using again Eq.~(\ref{shiftfinal}). The following remark is crucial:
to the order $d/R$ that we consider,  the Green's function ${\bar G}^{(S)}_{ij}(d)$ in Eqs.~(\ref{Gxxyy}--\ref{Gzz})   is no longer invariant under rotations around the $z$ axis. However ${\bar G}^{(S)}_{ij}(d)$ is still invariant under reflections of the $x$ and $y$ coordinates.  In order to take advantage of this reflection symmetry,  within each doublet $E_{l,|m|}^{(\rm plane)}$, $m \neq 0$  we replace the two states $|l, \pm m \rangle$  by the new states $|l, |m|, s \rangle$, with $s= \pm 1$, given by
\be
|l, |m|, \pm \rangle = \frac{1}{\sqrt{2}}(|l, m \rangle \pm (-1)^{|m|}\, |l, -m \rangle)\;,\;\;\;\;\;m=1,\dots, l
\ee
which possess definite parity under  independent  reflections of the coordinates $x$ and  $y$. For the $m=0$ singlets, we just set
\be
|l, 0, + \rangle \equiv |l, 0 \rangle\;.
\ee
Since
\be
{\hat R}_x |l, m \rangle = |l, -m \rangle \,,\label{Rx}
\ee
\be
{\hat R}_y |l, m \rangle = (-1)^m |l, -m \rangle \,,\label{Ry}
\ee
it is easy to verify  that the states $|l, |m|, \pm \rangle$ indeed have definite parity under reflections of $x$ and $y$:
\begin{align}
{\hat R}_x |l, |m|, \pm \rangle&=\pm\,(-1)^{|m|} \,  |l, |m|, \pm \rangle,\; m=0,\dots, l\\
{\hat R}_y |l, |m|, \pm \rangle&=\pm  |l, |m|, \pm \rangle,\;m=0,\dots, l \;.
\end{align}
Since to order $d/R$ the  Green's function is reflection-invariant, the CP interaction does not mix rotational states of different parity, 
and therefore  in the basis $|l, |m|, s \rangle$ we are allowed to  use the non-degenerate theory underlying Eq.~(\ref{shiftfinal})   to compute the {\it leading} curvature correction to the energy levels $E^{\rm (plane)}_{l,|m|}$. The matrix elements of ${\hat \mu}_i^2$ in the new basis are
\begin{align}
\langle l,|m|,s |\hat{\mu}_x^2| l, |m|,s \rangle&=\langle l,|m|,s | \hat{\mu}_y^2 | l, |m|,s \rangle&& \nonumber \\ 
&= \mu^2 \;\frac{l (l+1)+m^2-1}{4 l (l+1)-3}\;,\; |m| \neq 1\, ,
\end{align}
\begin{align}
\langle l,1,+ | \hat{\mu}_x^2| l, 1,+ \rangle&=\langle l,1,- | \hat{\mu}_y^2 | l, 1,- \rangle &&\nonumber \\ 
&=3\, \mu^2\;\frac{l (l+1)}{8 l (l+1)-6}\;, 
\end{align}
\begin{align}
\langle l,1,- | \hat{\mu}_x^2 | l, 1,- \rangle&=\langle l,1,+ | \hat{\mu}_y^2 | l, 1,+ \rangle \nonumber\\
&= \mu^2 \;\frac{l (l+1)}{8 l (l+1)-6}\;, 
\end{align}
and
\be
\langle l,|m|,s | \hat{\mu}_z^2 | l, |m|,s \rangle=\mu^2 \;\frac{2 l (l+1)-2 m^2-1}{4 l (l+1)-3}\;.
\ee
Using the above relations, the leading curvature correction $\Delta F^{\rm (curv)}_{l,|m|,s}$  to the rotational energy levels is found to be
\begin{widetext}
\be
\Delta F^{\rm (curv)}_{l,|m|,s}= {\cal E}  \left(\frac{d}{R_1}+\frac{d}{R_2} \right) \frac{ l (l+1)(11 + 5 \epsilon_{\rm st})+m^2 (\epsilon_{\rm st}-1)-4(2+\epsilon_{\rm st})}{4(\epsilon_{\rm st}+1)[4 l(l+1)-3]}\;,\;\;\;\;|m| \neq 1\;,\label{curv1}
\ee
\be
\Delta F^{\rm (curv)}_{l,1,\pm}= {\cal E}   \left\{\left(\frac{d}{R_1}+\frac{d}{R_2} \right)  \frac{ l (l+1)(11 + 5 \epsilon_{\rm st}) -3(3+\epsilon_{\rm st})}{4(\epsilon_{\rm st}+1) [4 l(l+1)-3]} \pm \left(\frac{d}{R_1}-\frac{d}{R_2} \right) \frac{ l (l+1)(1 + 3 \epsilon_{\rm st})  }{16(\epsilon_{\rm st}+1)[4 l(l+1)-3]}\right\}\;.\label{curv2}
\ee
\end{widetext}
We see that   for  $|m| >1$ surface  curvature   just determines an extra overall shift in the energy  of the  doublets $E^{\rm (plane)}_{l,|m|}$, without lifting their two-fold degeneracy. By contrast, the $|m|=1$ doublets split  into two distinct levels,  whose spacing is proportional to $(d/R_1-d/R_2)$ (see  Fig.~\ref{levelsS}).  The splitting of the $|m|=1$ rotational levels constitutes the characteristic signature of curvature effects on the CP interaction of the molecule with the surface.

\section{Structure of the rotational spectrum}

In a polar molecule, rotational transitions between adjacent rotational levels ($\Delta l = \pm 1$) are electric-dipole allowed~\cite{brown}. Let us consider as an example the emission lines corresponding to transitions from $l=1$ states to the rotational ground state $l=0$, i.e.  $l=1 \rightarrow 0$. When the molecule is far from the surface, these transitions correspond to a single spectral line of frequency $\nu_r=\omega_r/(2 \pi)$ (see Table I). As the molecule approaches  the surface, this line splits into several components. The precise number of lines depends on whether the surface is planar or curved. Let us consider first the case of  a planar surface. According to Eq.~(\ref{plane}), the free-space line $1 \rightarrow 0$ splits in two components corresponding to the transitions
$$\nu_1 :\;\;\;|1,\pm 1\rangle \rightarrow |0,0 \rangle\;,
\quad{\rm and}\quad \nu_2 :\;\;\; |1,0\rangle \rightarrow |0,0 \rangle\;.$$

Suppose  that we observe the molecule from a point along the $z$-axis, i.e. in a direction perpendicular to the planar surface. Since the $x$ and $y$ components of the dipole-moment operator $ \hat{\mu}_x$ and $\hat{\mu}_y$ do not couple two $m=0$ states, it follows that in the dipole approximation the line $\nu_2$ cannot be seen from this observation direction, and only the line $\nu_1$ is detected. When the observation line is instead in the plane of the surface, both lines are visibile, and is it easy to verify that the line $\nu_1$ is polarized in the plane of the surface, while the line $\nu_2$ is polarized along the normal direction  to the surface. According to Eq.~(\ref{plane}), the difference $\Delta \nu_{12}=\nu_1-\nu_2$ between the two lines is
\be
\Delta \nu_{12}=\nu_1-\nu_2= \;\frac{{\cal E}}{5\,h}\;,\label{delnuplane}
\ee
with ${\cal E}$ as defined in Eq.~(\ref{scale}).

For a {\it curved surface}, Eqs.~(\ref{curv1}--\ref{curv2}) indicate that the line $\nu_1$ of the planar surface splits into two components $\nu_1^{(\pm)}$ corresponding to the transitions  (see  Fig.~\ref{levelsS})
\begin{align}
\nu_1^{(+)}:\;\;\;|1,1,+\rangle \rightarrow |0,0,+\rangle\;,\nonumber\\  
\nu_1^{(-)}:\;\;\;|1,1,-\rangle \rightarrow |0,0,+\rangle\;.
\end{align}

According to Eq.~(\ref{curv2}) the difference $\Delta \nu_{\pm}$ between the frequencies $\nu_1^{(+)}$ and $\nu_1^{(-)}$ of these two lines is proportional to the difference in radii of curvature, as
\be
\Delta \nu_{\pm}=\nu_1^{(+)}-\nu_1^{(-)}= \frac{{\cal E}}{h } \left(\frac{d}{R_1}-\frac{d}{R_2} \right) \frac{1}{20} \;\frac{3\; \epsilon_{\rm st} +1 }{\epsilon_{\rm st}+1}\;.\label{delnucur}
\ee
In addition to the two lines $\nu_1^{(\pm)}$, we of course have a third line, corresponding to the line $\nu_2$ of the planar surface:
$$\nu_2:\;\;\;|1,0,+\rangle \rightarrow |0,0,+\rangle\;,$$
Thus, the single $l=1 \rightarrow 0$ line of free-space  splits (in general) into three lines, when the molecule is brought near a curved surface. 

Suppose again that we observe the molecule from a point along the $z$-axis.  Reasoning as before, we see that in the dipole approximation the line $\nu_2$ cannot be detected from this observation direction, and   only the  lines $\nu_1^{(+)}$ and $\nu_1^{(-)}$ are visible. Using  Eqs.~(\ref{Rx}) and (\ref{Ry}) it is easy to verify that  $\nu_1^{(+)}$ and $\nu_1^{(-)}$ are linearly polarized along the $x$ and the $y$ axis, respectively.

Similarly, it is possible to verify that when the observation direction is along the $x$-axis ($y$-axis),  the visible  lines are  $\nu_1^{(-)}$ ($\nu_1^{(+)}$) and  $\nu_2$; 
the former  linearly polarized in the $y$-direction ($x$-direction), and the latter along the $z$-axis.     
Up to small curvature corrections, the frequency differences $\Delta \nu_{12}^{(\pm)}=\nu_1^{(\pm)}-\nu_2$     coincide with the frequency difference $\Delta \nu_{12}$ for the planar surface in Eq.~(\ref{delnuplane}).  
By comparing Eq.~(\ref{delnucur}) with Eq.~(\ref{delnuplane}) we thus see that the curvature-induced splitting $\Delta \nu_{\pm}$, is suppressed by factor of order $d/R$, compared to the splittings $\Delta \nu_{12}^{(\pm)}$. From our perspective, though, the most interesting quantity to observe is   $\Delta \nu_{\pm}$ since it represents a pure curvature effect. 
Using Eq.~(\ref{scale}),   we  estimate the  magnitude of $\Delta \nu_{\pm}$ for a polar molecule with an electric dipole moment $\mu= 2 \times 10^{-29}$ Cm (see Table I), as

\begin{align}
\Delta \nu_{\pm}& \simeq &\frac{ 3\, \mu^2}{640 \, \pi \, \epsilon_0\, h \, d^3}\left(\frac{d}{R_1}-\frac{d}{R_2} \right)
\nonumber\\  
&=& 100\; {\rm kHz} \;\left(\frac{100\, {\rm nm}}{d}\right)^3 \;\left(\frac{d}{R_1}-\frac{d}{R_2} \right) \;.
\end{align}
Note that our derivation only assumes that $d/|R_1| \ll 1$ and $d/|R_2| \ll 1$. However it does not assume that $|R_1-R_2|/|R_1|\ll 1$.  In
particular, in the case of a cylindrical surface $R_1\to\infty$ and
$|R_1-R_2|/|R_1|=1$.
To determine if the frequency difference $\Delta \nu_{\pm}$ is potentially measurable, it is important to compare   $\Delta \nu_{\pm}$ with the typical width of rotational spectral lines.   Their natural width $\Delta \nu$ can be estimated by the simple formula ~\cite{brown}
 \be
{\Delta \nu}  =\frac{\nu^3 |\mu|^2}{3\, \epsilon_0  \, \hbar\, c^3} \;.
\ee  
For the molecules listed in Table I, the natural line width  ranges from a maximum of $4 \times 10^{-4}$ Hz for LiH to a minimum of $1.2 \times 10^{-10}$ Hz for NaRb, and is thus many orders of magnitude smaller than $\Delta \nu_{\pm}$, for reasonable values of the separation $d$, and of   $d/R$. Next we consider the thermal Doppler broadening, which for a gas of molecules in equilibrium at temperature $T$ is given by~\cite{brown}
\be
\Delta \nu=\frac{2 \nu}{c} \sqrt{\frac{2 N_A k_B T \log 2}{M}}= 7.15 \times 10^{-7}\,(T/M_{\rm r})^{1/2}\,\nu\,,
\ee
where
$N_A $ is Avogadro's number, $M$  and $M_{\rm r}$ are the mass and the relative molecular mass of the molecule, respectively. Using the above formula, we estimate that at room temperature $T=300$K, the Doppler broadening ranges from a maximum of 2 MHz for LiH, to a minimum of   5 kHz for NaRb and NaCs.  So, while for the light molecule LiH the large  thermal Doppler broadening prevents observation of the frequency shift $\Delta \nu_{\pm}$ even at cryogenic temperatures,   in the case of the heavier molecules listed in Table I the thermal Doppler broadening is favorably smaller than $\Delta \nu_{\pm}$ even at room temperature.

\begin{acknowledgments}
We thank M. Zwierlein  for valuable discussions.  This
research was supported by the NSF through grant No. DMR-12-06323 (MK),
and by the U. S. Department of Energy (DOE) under cooperative
research agreement \#DF-FC02-94ER40818 (RLJ).
\end{acknowledgments}


\begin{thebibliography}{200}

\bibitem{parse} V. A. Parsegian, {\it Van der Waals Forces} (Cambridge University Press, 2005).

\bibitem{london} F. London, Z. Phys. {\bf 63}, 245 (1930).

\bibitem{polder} H.B.G. Casimir and D. Polder, Phys. Rev. {\bf 73}, 360 (1948).


\bibitem{failache} H. Failache, S. Saltiel, M. Fichet, D. Bloch, and M. Ducloy, Phys. Rev. Lett. {\bf 83}, 5467 (1999).

\bibitem{bimo1} G. Bimonte, T. Emig and M. Kardar, Phys. Rev. D {\bf 90}, 081702 (2014).

\bibitem{bimo2} G. Bimonte, T. Emig and M. Kardar, Phys. Rev. D {\bf 92}, 025028 (2015).

\bibitem{haroche} V. Sandoghdar, C.I. Sukenik, E.A. Hinds and S. Haroche, Phys. Rev. Lett. {\bf 68}, 3432 (1992).

\bibitem{ducloy} A. Laliotis, T.P. de Silans, I. Maurin, M. Ducloy, and D. Bloch, Nature Comm. DOI: 10.1038/ncomms5364 (2014).

\bibitem{Buhmann2008} S. Y. Buhmann, M. R. Tarbutt, S. Scheel, and E. A. Hinds, Phys. Rev. A {\bf 78}, 052901 (2008).

\bibitem{Ellingsen2009} S. A. Ellingsen, S. Y. Buhmann, and S. Scheel, Phys. Rev. A {\bf 79}, 052903 (2009).

\bibitem{wylie} J.M. Wylie and J.E. Sipe, Phys. Rev. A {\bf 30}, 1185 (1984); {\it ibid.} {\bf 32}, 2030 (1985); 

\bibitem{laliotis} A. Laliotis and M. Ducloy, Phys. Rev. A{\bf 91}, 052506 (2015)

\bibitem{dalvit} D. Dalvit, P. Milonni, D. Roberts, and F. da Rosa Eds. {\it Casimir Physics, Lecture Notes in Physics, Vol. 834} (Springer-Verlag, Berlin Heidelberg 2011) 

\bibitem{brown} J. Brown and A. Carrington {\it Rotational Spectroscopy of Diatomic Molecules} (Cambridge University Press 2003)



\bibitem{jackson} J. D. Jackson {\it Classical Electrodynamics} (John Wiley \& Sons, New York 1999).

\bibitem{Ellingsen2010} S. A. Ellingsen, S. Y. Buhmann, and S. Scheel, Phys. Rev. Lett. {\bf 104}, 223003 (2010).

\bibitem{smythe} W. R. Smythe {\it Static and Dynamic Electricity} (McGraw-Hill Book Company, New York 1950)

\bibitem{fosco2}  C. D. Fosco, F. C. Lombardo, and F. D. Mazzitelli, Phys. Rev.D {\bf 84}, 105031 (2011).

\bibitem{bimonte3} G. Bimonte, T. Emig, R. L. Jaffe, and M. Kardar, EPL {\bf 97}, 50001 (2012).

\bibitem{bimonte4} G. Bimonte, T. Emig, and M. Kardar, Appl. Phys. Lett. {\bf 100}, 074110 (2012).


 




\end{thebibliography}
\end{document}